\documentclass[sigplan,10pt]{acmart}
\usepackage{algorithm}
\usepackage{algpseudocode}
\usepackage{booktabs}
\usepackage[normalem]{ulem}
\renewcommand\footnotetextcopyrightpermission[1]{}

\AtBeginDocument{%
  }

\setcopyright{none}
\acmConference[ATC '26]{The 2026 ACM SIGOPS Annual Technical Conference}{November 15--18, 2026}{Shatin, Hong Kong}
\acmBooktitle{Proceedings of the 2026 ACM SIGOPS Annual Technical Conference (ATC '26), November 15--18, 2026, Shatin, Hong Kong}
\acmYear{2026}
\copyrightyear{2026}
\acmDOI{}
\acmISBN{}
\begin{document}



\title{Coordinated Scheduling for MoE LLM Serving}

\author{Yifan Sun}
\affiliation{
  \institution{The University of Melbourne}
    \city{Melbourne}
  \country{Australia}
}

\author{Zhexiang Zhang}
\affiliation{
  \institution{The University of Melbourne}
    \city{Melbourne}
  \country{Australia}
}

\author{Jiantong Jiang}
\affiliation{
  \institution{The University of Melbourne}
    \city{Melbourne}
  \country{Australia}
}

\author{Gholamreza Haffari}
\affiliation{
  \institution{Monash University}
    \city{Melbourne}
  \country{Australia}
}

\author{Minxian Xu}
\affiliation{
  \institution{Shenzhen Institutes of Advanced Technology, Chinese Academy of Sciences}
      \city{Shenzhen}
  \country{China}
}

\author{Feng Liu}
\affiliation{
  \institution{The University of Melbourne}
    \city{Melbourne}
  \country{Australia}
}

\author{Rajkumar Buyya}
\affiliation{
  \institution{The University of Melbourne}
    \city{Melbourne}
  \country{Australia}
}

\author{Adel N. Toosi}
\affiliation{
  \institution{The University of Melbourne}
    \city{Melbourne}
  \country{Australia}
}

\begin{abstract}
Serving Mixture-of-Experts (MoE) large language models (LLMs) is challenging because dynamic request workloads interact with sparse expert routing, creating both data-parallel (DP) engine imbalance and expert-level hotspots. Existing LLM serving systems typically make these decisions in isolation: frontend schedulers route requests using coarse request counters, while backend expert balancers rely mainly on aggregate expert activation counts. This separation prevents the serving system from reacting to fine-grained engine pressure, backend MoE pressure, and source-dependent expert traffic.
To address this gap, we propose Gimbal, a coordinated cross-level scheduling system for efficient MoE-based LLM serving. First, Gimbal presents a fine-grained DP-engine scheduler that uses online backend pressure signals, including key-value (KV) cache usage, remaining prefill work, queue backlog, and MoE expert pressure, to dispatch requests away from overloaded engines. Inside each engine, Gimbal further applies a lightweight prefill-aware queue ordering policy with aging to reduce head-of-line blocking without output-length prediction. Second, Gimbal extends expert load balancing with online source-DP-to-expert routing statistics and uses a heuristic guided by a mixed-integer nonlinear program (MINLP) to place experts while jointly considering expert load, source-aware communication, and migration stability. Our evaluation shows that Gimbal reduces average Time To First Token (TTFT) by 42.9\% and average Time Per Output Token (TPOT) by 33.3\% compared with the state-of-the-art serving system such as vLLM, while improving high-load request throughput by 3.0\%.
\end{abstract}

\maketitle
\pagestyle{empty}
\thispagestyle{empty}

\section{Introduction}

Large Language Models (LLMs) have rapidly advanced from generative pretraining to few-shot and conversational systems~\cite{radford2018improving,gpt2,brown2020language,openai2022chatgpt}.
Early advances in LLMs were largely driven by scaling dense Transformer-based models~\cite{vaswani2023attentionneed}, 
 but such scaling also increases
demands on computational resources, inference latency, and deployment cost. Therefore, efficiently serving such models in real-world applications has become a critical problem that warrants immediate attention~\cite{jiang2026towards}.
To address this concern, sparse architectures, especially Mixture-of-Experts (MoE) models~\cite{shazeer2017outrageously,lepikhin2021gshard,mix-of-experts,du2022glam,jiang2024mixtralexperts}, have reshaped large-model scaling. DeepSeek-series models~\cite{deepseekai2025deepseekr1incentivizingreasoningcapability,deepseekai2026deepseekv4,deepseekai2024deepseekv3technicalreport,deepseekv2} show that sparse MoE models can achieve competitive capability while substantially reducing activated computation. By activating only a small subset of experts per token, MoE models can scale model capacity while keeping per-token inference cost manageable.

Although MoE architectures hold great promise, they introduce new challenges for system design and serving. MoE inference systems typically combine \textit{data parallelism} (DP), \textit{expert parallelism} (EP), and often \textit{tensor parallelism} (TP). DP replicates inference engines to improve throughput, EP distributes experts across GPUs or nodes to serve sparse expert computation, and TP partitions dense model computation across GPUs~\cite{lepikhin2021gshard,rajbhandari2022deepspeedmoe,hwang2023tutel,deepseekai2024deepseekv3technicalreport}. This layered parallelism improves scalability, but it also makes scheduling decisions tightly coupled across frontend request dispatching and backend expert execution. Current LLM serving systems, however, largely inherit scheduling policies designed for dense models~\cite{yu2022orca,vllm,zheng2024sglang,agrawal2023sarathi,agrawal2024sarathiserve}, with these decisions still made mostly in isolation: frontend schedulers typically rely on coarse request-level metrics, while backend expert balancers depend primarily on aggregate expert activation counts. In particular, DP-engine selection often relies on Round-Robin or request-count-based dispatching. These policies implicitly assume that each request has a similar execution cost, but this assumption does not hold in real workloads, where prompt lengths can vary significantly~\cite{burstgpt,zheng2023lmsyschat1m}.
In MoE serving, DP-level imbalance, sparse expert imbalance, and cross-DP communication are tightly coupled rather than independent.
A request assigned to a DP engine not only consumes local prefill computation and KV-cache resources; its tokens are also routed to a small subset of experts, creating expert-level hotspots and
all-to-all communication pressure inside MoE layers. Since the assigned DP engine becomes the source of this token traffic, DP-engine dispatch decides where MoE all-to-all communication originates, while expert placement decides whether the selected experts are local or remote relative to that source.
In fact, request-count-based or Round-Robin DP scheduling cannot capture token-level load or backend MoE pressure. At the same time, EPLB-style mechanisms~\cite{deepseekai2024deepseekv3technicalreport,deepseekai_eplb}
can rebalance aggregate expert computation but do not explicitly optimize source-dependent cross-DP communication. Therefore, effective MoE serving requires coordinated cross-level scheduling rather than separately improving the DP scheduler or the expert load balancer in isolation.

An ideal MoE scheduler should satisfy two requirements. First, it needs fine-grained load
awareness, capturing runtime pressure such as prefill work, waiting-queue backlog, KV-cache
pressure, and backend expert pressure, rather than relying only on request counts. Second,
it needs coordinated expert placement, jointly considering expert load, source-aware
communication cost, and migration overhead so that expert placement matches the observed
request-to-expert traffic without introducing excessive movement.

To address these challenges, we propose Gimbal, a coordinated cross-level scheduling
framework for MoE-based LLM serving. Gimbal first introduces fine-grained DP-engine
scheduling, which selects DP engines based on running and waiting prefill tokens, KV-cache
usage, and backend MoE expert pressure. Within each engine, Gimbal further uses a
lightweight Shortest-Job-First (SJF)-style queue ordering with aging to reduce head-of-line blocking caused by
large prefill requests. Gimbal then incorporates runtime routing information into MoE expert
placement by collecting source-DP-to-expert activation statistics online and jointly
considering expert load, source-aware communication cost, and migration overhead.
  
Overall, Gimbal closes the loop between DP-engine scheduling and MoE expert placement. The DP layer mitigates request-side imbalance before it propagates into the backend, while the MoE layer performs dynamic, source-aware expert migration and placement to improve load balance, reduce expert hotspots, and lower communication cost. Meanwhile, backend signals such as expert pressure are fed back into frontend request scheduling decisions, enabling Gimbal to coordinate request dispatching and expert rebalancing during serving.
Experimental results show that this coordinated cross-level design improves serving latency
without sacrificing throughput. Across all request rates and workload distributions, Gimbal reduces average Time To First Token (TTFT) by 42.9\% and average Time Per Output Token
(TPOT) by 33.3\% compared with vLLM, while improving high-load request throughput by 3.0\%.
It also reduces P99 TTFT by 44.3\%. Ablation study results further show that coordinating DP
scheduling with source-aware expert placement outperforms simply enabling the two
optimizations independently.

\section{Background and Motivation}

\subsection{MoE LLM Serving Stack}
Modern LLM inference follows an autoregressive process with two main phases: (i) prefill, which processes the input prompt, and (ii) decode, which generates output tokens step by step. Serving systems cache key/value tensors in the KV cache to avoid recomputing previous context, making prompt length, remaining prefill work, and KV-cache usage important factors in request cost. To serve models at high throughput, systems such as vLLM~\cite{vllm}, SGlang~\cite{zheng2024sglang}, TGI~\cite{tgi}, and Preble~\cite{srivatsa2024preble} typically use a layered architecture with a frontend router and multiple backend inference engines. In DP serving, the router dispatches requests across replicated engines, while each replica batches its assigned requests for prefill and decode execution and maintains the corresponding per-request KV cache.

Mixture-of-Experts (MoE) models add another layer of complexity to this serving stack. Instead of activating all feedforward parameters, an MoE layer uses a lightweight router to select top-$k$ experts for each token. These experts are often distributed across GPUs through EP~\cite{lepikhin2021gshard,rajbhandari2022deepspeedmoe,hwang2023tutel,deepseekai2024deepseekv3technicalreport}. During MoE execution, tokens are dispatched to the GPUs that host their selected experts, and the computed expert outputs are then combined and returned, typically through all-to-all communication. As a result, MoE serving combines DP-engine request dispatching with EP-level expert routing and communication. This layered design improves scalability, but it also introduces new coordination challenges between the DP scheduler, which determines where requests enter the system, and the MoE layers, which determine how tokens are routed across experts and devices.

\subsection{Observations}


\subsubsection{Coarse request scheduling causes inter-engine imbalance}

\begin{figure}[t]
    \centering
    \includegraphics[width=\columnwidth]{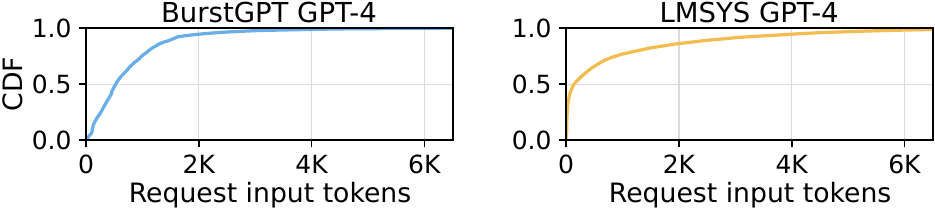}
    \caption{Empirical CDFs of GPT-4 request input lengths in BurstGPT~\cite{burstgpt} and LMSYS-Chat-1M~\cite{zheng2023lmsyschat1m}. Real user requests exhibit substantial length heterogeneity across datasets.} 
    \label{fig:gpt4-request-length-distribution}
    \includegraphics[width=\columnwidth]{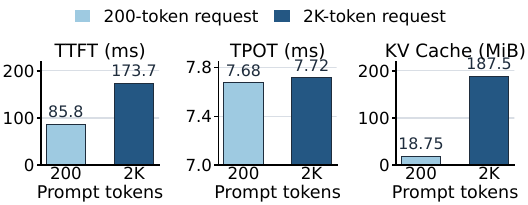}
    \caption{Single-request cost using Qwen3-80B-INT4 on an A100 GPU. The same number of requests can result in completely different engine loads.}
                
    \label{fig:request-granularity}
\end{figure}

Request-level scheduling treats requests as roughly equivalent work units, obscuring substantial differences in computation, communication, and memory costs across requests. In practice, a request's execution pressure depends on fine-grained factors, including prompt length, remaining prefill tokens, decode work, and total KV-cache usage. Figure~\ref{fig:gpt4-request-length-distribution} shows that real GPT-4 workloads exhibit highly skewed input lengths: while many requests are short, both BurstGPT and LMSYS-Chat-1M contain long-tail requests with thousands of input tokens. This length heterogeneity translates into different prefill computation and KV-cache pressure, making request count a coarse engine-load indicator.

Figure~\ref{fig:request-granularity} further illustrates this gap with two single-request examples. Although both cases contain exactly one request, their serving pressure differs significantly. The 2K-token request consumes 187.5MiB of KV cache, while the 200-token request consumes only 18.75~MiB. Their TTFT and TPOT also differ accordingly. This shows that two DP engines with the same number of active requests may still experience very different computation, memory, and latency pressure. Therefore, DP-engine scheduling should use fine-grained runtime states rather than request-level counts.

\vspace{1.5mm}
\noindent\fbox{%
\begin{minipage}{0.94\columnwidth}
\emph{\textbf{Fine-grained DP scheduling:} DP-engine scheduling should move from request-level counters to token-level runtime pressure.}
\end{minipage}}

\subsubsection{Sparse routing causes intra-model expert and communication imbalance}

\begin{figure}[t]
    \includegraphics[width=\columnwidth]{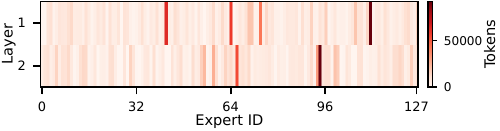}
    \caption{Expert heatmap showing activation hotspots in certain MoE layers.}
    \label{fig:expert-hotspots}
    \includegraphics[width=\columnwidth]{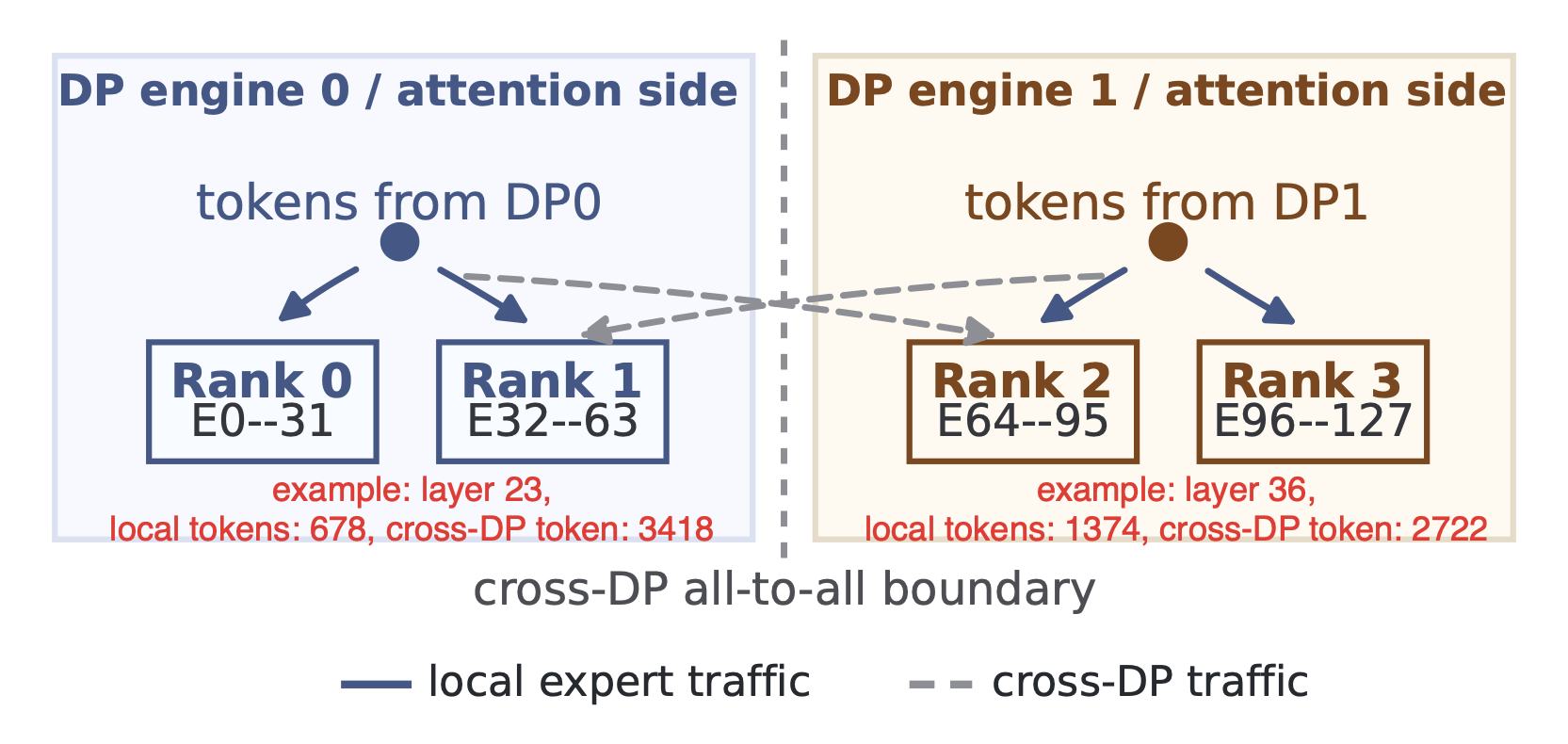}
    \caption{Expert placement determines whether MoE traffic stays local or crosses DP groups. Profiling-window examples show that 83.4\% of Layer-23 traffic from DP0 and 66.5\% of Layer-36 traffic from DP1 are routed to remote DP groups under the current placement.}
    \label{fig:source-expert-comm}
\end{figure}

MoE models activate only a small subset of experts for each token~\cite{shazeer2017outrageously,mix-of-experts,deepseekv2}, so token traffic is not uniformly distributed across experts. A small number of experts may receive disproportionately high activation load, while other experts are rarely used. Figure~\ref{fig:expert-hotspots} shows expert activation results from serving 1,000 requests on Qwen3-80B-INT4 MoE. In some layers, the hottest experts receive far more tokens than the layer average. When EP is enabled, if several hotspot experts are placed on the same GPU, this expert-load imbalance leads to uneven computation across devices and reduces inference efficiency.

Meanwhile, when EP is used together with DP and TP, each request is assigned to one DP engine, and the tensor-parallel GPUs in that DP group jointly execute the attention and dense layers. When execution reaches an MoE layer, token hidden states are dispatched to the GPUs that host the selected experts, and the results are later merged through all-to-all communication. Prior work has shown that MoE routing is not purely random; expert activations often exhibit exploitable structure, such as inter-layer expert affinity and stable expert traffic skew~\cite{go2025moetuner,yao2024exploiting}. As shown in Figure~\ref{fig:source-expert-comm}, this effect appears in our profiling traces. Under the current placement, 83.4\% of Layer-23 traffic from DP0 and 66.5\% of Layer-36 traffic from DP1 are routed to experts located in remote DP groups. This source-dependent traffic skew shows that expert placement can directly affect cross-DP all-to-all communication, and motivates using source-aware routing statistics when rebalancing experts.


However, expert migration is not free. In our H100 NVLink testbed, the first all-layer expert rearrangement takes 1.08~s on average, and subsequent online rearrangements take 0.72~s on average. This cost is small compared with long-running serving windows, but it is still large enough that expert placement should avoid unnecessary movement. Therefore, source-aware expert placement must balance three goals: reducing expert hotspots, reducing cross-DP communication, and limiting migration overhead.

\vspace{1.5mm}
\noindent\fbox{%
\begin{minipage}{0.94\columnwidth}
\emph{\textbf{Source-aware expert placement:} Dynamic expert placement should jointly consider expert load, observed source-DP-to-expert traffic skew, and migration cost, placing hot experts closer to their dominant observed sources when such skew exists, while preserving load balance and avoiding unnecessary movement.}
\end{minipage}}

\subsubsection{Inter-engine and intra-model imbalance are coupled}
Inter-engine imbalance and intra-model imbalance are not independent in MoE serving. DP-engine dispatch determines where a request enters the serving stack, while sparse routing determines which experts its tokens visit inside MoE layers. Although the DP scheduler does not control the router's expert choices, it determines the DP source from which routed tokens enter the all-to-all communication path. As a result, the same expert traffic can incur different communication pressure depending on the request's DP source and the current expert placement.

This coupling is further amplified by DP+TP+EP deployments. The attention and dense-layer computation of a DP engine often share physical GPU ranks with MoE expert computation and all-to-all communication. When some ranks experience high expert load or communication pressure, the effective capacity of the co-located DP engine can decrease. Therefore, backend expert pressure should not be treated as an isolated MoE-layer signal; it should be fed back to DP-engine scheduling so that new requests avoid engines whose underlying ranks are already under high MoE pressure.

\vspace{1.5mm}
\noindent\fbox{%
\begin{minipage}{0.94\columnwidth}
\emph{\textbf{Expert-pressure feedback:} Backend expert pressure should be exposed to DP-engine scheduling because it affects effective engine capacity.}
\end{minipage}}

\subsection{Limitations of Current Systems}
Current LLM serving systems mostly address these issues through component-specific optimizations. 

At the DP-engine and request-routing level, systems such as vLLM~\cite{vllm}, SGlang~\cite{zheng2024sglang}, FastServe~\cite{wu2024fastdistributedinferenceserving}, SeaLLM~\cite{zhao2025seallmserviceawarelatencyoptimizedresource}, BROS~\cite{borui2025efficientllmservinghybrid}, and Preble~\cite{srivatsa2025preble} improve dispatching with request-count-based routing, preemption, priority scheduling, phase-aware routing, or prefix-aware placement. Other systems further reorder requests inside each engine using length-aware, priority-based, or prediction-based policies~\cite{fu2024efficient,qiu2024efficientinteractivellmserving,zhao2025seallmserviceawarelatencyoptimizedresource,wu2024fastdistributedinferenceserving}. These techniques improve request scheduling, but they mainly operate on request-level or phase-level signals. They do not directly expose fine-grained online engine pressure, such as remaining prefill work, waiting-token pressure, KV-cache usage, or backend MoE pressure. As a result, engines that appear balanced by request count may still be imbalanced in computation, memory use, or expert-side contention.
At the MoE layer, EPLB-style mechanisms~\cite{deepseekai2024deepseekv3technicalreport,deepseekai_eplb} collect aggregate expert activation counts and rebalance experts across EP ranks, while MoETuner~\cite{go2025moetuner} uses offline expert affinity to guide placement. These approaches reduce expert hotspots or preserve affinity, but they optimize expert placement largely independently from DP-engine dispatch. Aggregate expert load also does not distinguish which DP source generates the traffic, so a placement can balance total expert load while still increasing cross-DP all-to-all communication. Sem-MoE~\cite{li2026semantic} combines static expert placement with semantic-aware routing, but its offline profiling and prediction can become stale as arrivals, prompt lengths, routing patterns, and expert hotspots change during serving.

These limitations motivate coordinated cross-level scheduling rather than independent optimization of components. A serving system should use fine-grained runtime states to balance DP-engine pressure, use source-aware routing statistics to guide expert placement, and feed backend expert pressure back to DP-engine scheduling. Gimbal is designed around this feedback loop, coordinating DP-engine dispatch and source-aware expert placement during online serving.

\section{Overall System Design}
\begin{figure}[h]
    \centering
    \includegraphics[width=\columnwidth]{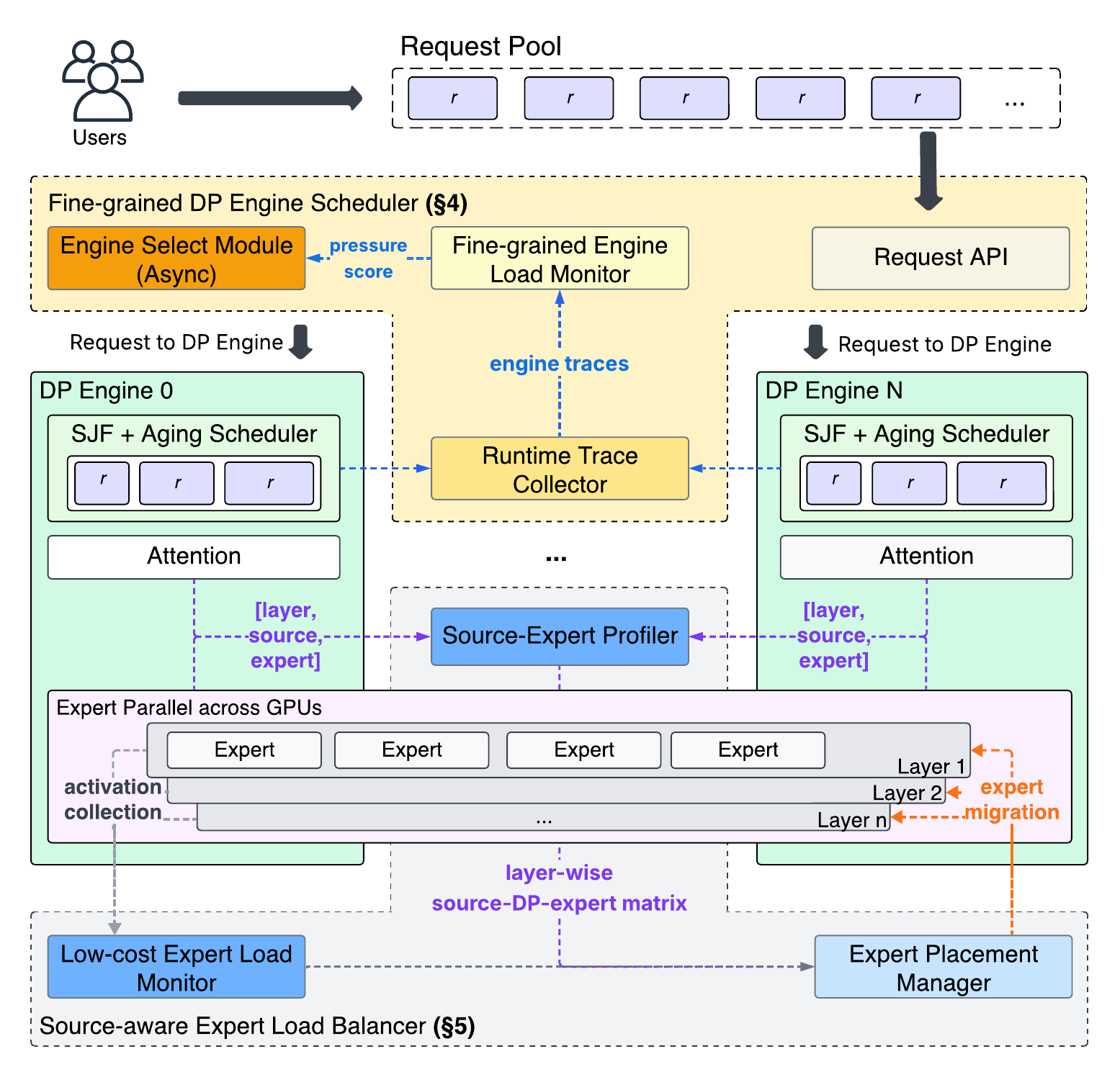}
            \vspace{-0.7cm}
    \caption{Gimbal system architecture. Gimbal coordinates a fine-grained DP-engine scheduler and a source-aware expert load balancer in MoE serving. } 

    \label{fig:gimbal-architecture}
\end{figure}

This section presents the overall design of Gimbal. As shown in Figure~\ref{fig:gimbal-architecture}, Gimbal coordinates two main components during MoE serving: a fine-grained DP-engine scheduler (\S\ref{sec:dp-engine-scheduling}) and a source-aware expert load balancer (\S\ref{sec:source-aware-expert-lb}). The dark arrows denote the request path, the blue dashed arrows denote runtime engine traces used by the DP scheduler, the purple dashed arrows denote source-expert profiling statistics, the gray dashed arrows denote aggregate expert-load statistics, and the orange dashed arrows denote expert migration actions.

Incoming requests first enter the request pool and are submitted through the Request API. Instead of assigning requests according to request counts, Gimbal uses the Engine Select Module to choose a DP engine based on fine-grained runtime pressure. This decision is made asynchronously from the request path: backend DP engines periodically report engine traces to the Runtime Trace Collector, including remaining prefill tokens, waiting prefill tokens, KV-cache usage, and backend MoE pressure. The Fine-grained Engine Load Monitor converts these traces into a comparable pressure score for each DP engine, and the Engine Select Module dispatches new requests to the engine with the lowest pressure. Inside each DP engine, Gimbal further applies a lightweight SJF-style queue ordering with aging as a local optimization to reduce head-of-line blocking.

At the MoE execution layer, Gimbal collects two types of routing statistics during inference. First, the Low-cost Expert Load Monitor records aggregate expert activation counts, which capture how much computation load each expert receives. Second, the Source-Expert Profiler records a layer-wise source-expert matrix indexed by layer, DP source, and expert. This matrix captures which DP sources generate traffic to which experts, providing a lightweight online routing profile for source-aware placement. These statistics are collected along the normal MoE dispatch path, so they do not require a separate profiling pass or offline warm-up.

The Source-aware Expert Load Balancer consumes both aggregate expert-load statistics and the layer-wise source-expert matrix. The Expert Placement Manager uses aggregate expert load to avoid computation hotspots, uses source-expert traffic to estimate cross-DP communication cost, and considers migration overhead before applying a new placement. The resulting placement is applied to the expert-parallel region through online expert migration. In this way, Gimbal does not optimize DP-engine scheduling and MoE expert placement independently. Instead, backend expert pressure is fed back to the DP scheduler, while source-aware routing statistics guide expert placement, forming a coordinated feedback loop between request dispatch and MoE execution.

\section{Fine-grained DP-engine Scheduling}
\label{sec:dp-engine-scheduling}

Gimbal treats DP-engine scheduling as a pressure-aware admission-control problem. Instead of simply using request count as a proxy for engine load, Gimbal selects the target DP engine according to fine-grained runtime states reported online by backend engines. The scheduler has two levels. At the inter-engine level, Gimbal assigns each new request to a suitable DP engine according to the online pressure of different engines. At the intra-engine level, each engine further applies a lightweight SJF-style ordering with aging to reduce head-of-line blocking in the local queue.

\subsection{Runtime Trace Collection}

Each backend DP engine periodically and asynchronously reports a compact runtime trace to the global scheduler. This trace contains four types of state: 1) remaining prefill tokens from running requests, 2) waiting prefill tokens in the local waiting queue, 3) KV-cache usage, and 4) backend MoE expert pressure. These signals reflect different sources of engine pressure: remaining prefill tokens reflect ongoing computation pressure, waiting prefill tokens reflect local queueing pressure, KV-cache usage reflects memory pressure, and MoE expert pressure reflects backend contention from the expert-parallel execution path.

We use remaining prefill tokens instead of the original prompt length because modern LLM serving systems rely on continuous batching, and prefill execution may be split across multiple scheduling iterations under chunked prefill or token-budget constraints. Therefore, a request's prompt length at admission time does not always represent the amount of prefill work that still remains on an engine. Gimbal tracks unfinished prefill work from running requests and separates it from waiting prefill work in the local queue, allowing the scheduler to distinguish active computation pressure from queueing pressure.

Gimbal also includes backend MoE expert pressure in the runtime trace. When reporting runtime states, each DP engine collects the recent number of tokens processed by experts on its associated EP ranks. This signal captures the pressure introduced by MoE execution. In common DP+TP+EP deployments, attention/dense computation and expert computation share physical GPU resources. Therefore, high expert load or MoE communication on a rank can reduce the effective capacity available to the co-located DP engine. Feeding this pressure back to the DP scheduler helps avoid assigning more requests to engines whose underlying ranks are already under high MoE pressure.

This MoE pressure signal also serves as a bridge between the upper-level DP scheduler and the lower-level expert placement module. Since Gimbal performs online expert migration, expert placement decisions can change the distribution of expert load across EP ranks over time. If the DP scheduler is unaware of these changes, it may continue assigning requests to co-located engines whose backend pressure has increased after rebalancing. By feeding backend MoE pressure into DP-engine scheduling, Gimbal closes this feedback loop: expert placement reduces expert-side hotspots, while the DP scheduler avoids amplifying newly observed backend pressure through request dispatch.

Runtime trace collection is asynchronous with request admission. Request submission does not wait for a fresh trace; instead, the scheduler uses the latest available trace from each engine. Since each trace only contains a small set of scalar counters, the collection overhead is low. If traces are temporarily incomplete, Gimbal falls back to ordered dispatch, ensuring that request admission remains reliable before all backend states become available.
\subsection{Pressure-Aware Engine Selection}
Gimbal uses the collected traces to compute a comparable pressure score for each DP engine. The design follows two principles.
First, KV-cache pressure is prioritized because KV cache is a hard resource constraint in serving. If one engine already has high KV usage and the KV gap across engines is clear, sending new requests to the engine with high KV usage can further worsen memory pressure and reduce scheduling flexibility. When an inference system exhausts KV cache, it may also trigger more severe problems such as preemption and recomputation, which can significantly degrade performance. Therefore, in this case, Gimbal directly selects the engine with lower KV usage.
Second, when KV cache does not create a clear cross-engine imbalance, Gimbal falls back to a score-based comparison of overall engine pressure. This score jointly considers remaining prefill work, waiting-token pressure, lightweight compensation for recently dispatched requests, KV-cache penalty, and MoE expert-pressure penalty. This keeps the common scheduling path simple: each engine is compressed into a comparable pressure value, while the value still captures computation, queueing, memory, and backend MoE pressure.

Gimbal also uses a lightweight compensation term because runtime traces are updated asynchronously. If the scheduler only trusts the latest trace, it may repeatedly select the same engine before the next trace update arrives, causing transient load imbalance. Therefore, after each request assignment, Gimbal immediately adds a temporary pressure estimate to the selected engine, allowing the scheduler to account for recently assigned requests before backend traces are refreshed.

\subsection{Scheduling Workflow}
\begin{algorithm}[t]
    \caption{Global DP Engine Scheduling}
    \label{alg:global-dp-engine-scheduler}
    \begin{algorithmic}[1]
        \Statex \textbf{Input:} incoming request \(r\), DP engines \(\mathcal{E}\),
        \Statex \hspace{\algorithmicindent} backend traces \(\mathcal{T}=\{kv_i, pre_i^{rem}, wait_i, moe_i\}_{e_i\in\mathcal{E}}\),
        \Statex \hspace{\algorithmicindent} compensation load \(comp_i\)
        \Statex \textbf{Output:} selected DP engine \(e^*\)
        \If{\(\mathcal{T}\) is incomplete}
            \State \Return \(\textsc{Next}(\mathcal{E})\)
        \Else
            \State Load \(\{kv_i, pre_i^{rem}, wait_i, moe_i\}\) from \(\mathcal{T}\)
        \EndIf
        \State \(i_{min} \gets \arg\min_i kv_i\), \(i_{max} \gets \arg\max_i kv_i\)
        \If{\(\textsc{HighKV}(kv_{i_{max}})\) and \(\textsc{LargeGap}(kv_{i_{max}}-kv_{i_{min}})\)}
            \State \Return \(e_{i_{min}}\)
        \EndIf
        \For{each engine \(e_i \in \mathcal{E}\)}
            \State \(score_i \gets pre_i^{rem} + wait_i + comp_i + P_{kv}(kv_i) + P_{moe}(moe_i)\)
        \EndFor
        \State \(\Delta_s \gets \max_i score_i - \min_i score_i\)
        \If{\(\textsc{Close}(\Delta_s, r)\)}
            \State \Return \(\textsc{Next}(\mathcal{E})\)
        \EndIf
        \State \Return \(e_{\arg\min_i(score_i, kv_i, i)}\)
    \end{algorithmic}
\end{algorithm}
Algorithm~\ref{alg:global-dp-engine-scheduler} summarizes the inter-engine scheduling workflow. For each incoming request, Gimbal first checks whether backend traces are available. If traces are incomplete, it uses ordered dispatch as a fallback. Otherwise, the scheduler reads the latest KV usage, remaining prefill tokens, waiting-token pressure, and MoE pressure from the trace table.
The scheduler then checks KV-cache pressure. If the maximum KV usage is high and the KV-usage gap across engines is clear, Gimbal assigns the request to the engine with the lowest KV usage. Otherwise, it computes the pressure score for each engine and selects the engine with the lowest score. When the scores of different engines are close, Gimbal does not overreact to small numerical differences and instead falls back to ordered dispatch. This reduces scheduling oscillation caused by trace noise while still allowing the scheduler to react to clear pressure imbalance. This score-based workflow keeps request admission lightweight: KV cache is prioritized only when it becomes a clear bottleneck, while compute pressure, queue pressure, memory pressure, and backend MoE pressure are otherwise combined into a unified decision signal.

\subsection{Lightweight Intra-Engine Ordering}
After a request is assigned to a DP engine, it enters the local waiting queue of that engine. Gimbal uses a lightweight SJF-style ordering with aging inside each engine. This local policy can further reduce local head-of-line blocking. Since prefill length is known when a request arrives and has a strong impact on early computation cost, Gimbal uses prefill token count as a simple job-size estimate.
Algorithm~\ref{alg:SJF with Aging} shows this local ordering rule. Short prefill requests are scheduled earlier, approximating shortest-job-first behavior without predicting output length. To avoid starving long requests, Gimbal promotes requests whose waiting time exceeds the aging threshold to high priority. This keeps the policy lightweight and predictable while improving local queue behavior under workloads with heterogeneous prompt lengths.

\begin{algorithm}[t]
    \caption{Request Scheduler with SJF and Aging}
    \label{alg:SJF with Aging}
    \begin{algorithmic}[1]
        \Statex \textbf{Input:} \(waiting\_queue\), \(time_{now}\), threshold \(\theta_{age}\)
        \Statex \textbf{Output:} re-ordered \(waiting\_queue'\)
        \For{each request \( r \in waiting\_queue \)}
            \State waiting time \( w_r = time_{now} - r.arrival\_time \)
            \If{\( w_r \ge \theta_{age} \)}
                \State Assign \textbf{high priority} to \( r \) 
            \Else
                \State Assign priority based on request's prefill length \( r.prompt \)
            \EndIf
        \EndFor
        \State Sort \( waiting\_queue\) by priority ascending
        \State \Return \( waiting\_queue' \)
    \end{algorithmic}
\end{algorithm}

\section{Source-aware Expert Load Balancing}
\label{sec:source-aware-expert-lb}

The second major component of Gimbal is a source-aware expert load balancer. Its goal is to reduce both expert-side computation imbalance and source-dependent all-to-all communication. Existing EPLB-style~\cite{deepseekai_eplb} mechanisms mainly balance experts according to aggregate expert activation counts, which is effective for spreading expert computation across EP ranks. However, aggregate expert load alone does not reveal where the expert traffic comes from. As a result, a placement may balance total expert load while still sending many tokens across DP groups.

Prior work~\cite{yao2024exploiting} has shown that MoE routing often exhibits exploitable expert affinity: tokens that activate certain experts in one layer may frequently activate related experts in nearby layers, and colocating affinity-related experts on the same EP rank can improve locality and reduce cross-rank MoE communication~\cite{yao2024exploiting,go2025moetuner}. Gimbal builds on the same locality principle for online MoE serving. Instead of collecting full pairwise expert affinity online, which can be expensive across layers, requests, and experts, Gimbal uses a lightweight source-aware approximation: a layer-wise source-DP-to-expert activation matrix. For each MoE layer, the matrix records how many tokens from each DP source are routed to each expert. By stacking these matrices across layers, Gimbal obtains a lightweight online routing profile that captures source-conditioned expert traffic skew during serving. This profile lets Gimbal exploit observed source-expert traffic structure when possible, while still preserving aggregate expert-load balance and limiting migration overhead. 
Gimbal does not modify the MoE router or assume fixed expert placement. Instead, it observes expert traffic over profiling windows and updates expert placement accordingly. As shown in Figure~\ref{fig:gimbal-architecture}, the Source-aware Expert Load Balancer contains three logical components: 1) the Low-cost Expert Load Monitor, 2) the Source-Expert Profiler, and 3) the Expert Placement Manager. The first two collect runtime placement signals, while the Expert Placement Manager makes migration decisions based on expert load, source-aware communication cost, and migration overhead.

\subsection{Online Expert Traffic Profiling}

Gimbal collects two types of expert-side runtime statistics. First, the Low-cost Expert Load Monitor records aggregate expert activation counts. We denote this matrix as \(B_{l,e}\), where \(B_{l,e}\) is the number of tokens routed to expert \(e\) in layer \(l\) during a profiling window. This signal captures the computation load of each expert and is used to avoid placing an excessive number of hot experts on the same EP rank.

Second, the Source-Expert Profiler records a layer-wise source-DP-to-expert activation matrix. We denote this matrix as \(A_{l,s,e}\), where \(A_{l,s,e}\) is the number of tokens from DP source \(s\) that are routed to expert \(e\) in layer \(l\). This matrix captures the source-aware routing profile of MoE execution. Compared with aggregate expert load, \(A\) provides one more dimension: it reveals not only which experts are hot, but also which DP sources generate the traffic to those experts. Table~\ref{tab:source-expert-matrix-example} gives a concrete example of these two statistics.

Both statistics are collected along the normal MoE dispatch path. The aggregate expert-load matrix can reuse the expert activation counting path used by EPLB-style mechanisms. The source-aware matrix adds the DP-source dimension to this counting process and records \([layer, source, expert]\) statistics during token dispatch. This avoids a separate profiling pass and keeps the collection lightweight enough for online serving. To further reduce collection overhead, Gimbal also adds a fast path for source-aware statistics collection and implements a dedicated Triton kernel~\cite{tillet2019triton}, which we describe in the implementation section.

\begin{table}[h]
\centering
\caption{Example slice of Gimbal's collected expert statistics.}
\label{tab:source-expert-matrix-example}
\begin{tabular}{c c r r r}
\hline
Layer & Expert & \(B_{l,e}\) & \(A_{l,0,e}\) & \(A_{l,1,e}\) \\
\hline
29 & 18  & 104526 & 478 & 490 \\
29 & 71  & 51196  & 164 & 128 \\
29 & 80  & 92608  & 464 & 466 \\
29 & 82  & 70630  & 384 & 400 \\
\hline
\end{tabular}
\end{table}

\subsection{Source-aware Placement Objective}

The Expert Placement Manager uses the collected matrices to decide where experts should be placed. The placement objective has three goals. First, it should balance expert computation load across EP ranks. Second, it should reduce source-aware communication by placing frequently accessed experts closer to their dominant DP sources when possible. Third, it should limit migration overhead, since moving experts during online serving is not free.

Let \(x_{l,e,g} \in \{0,1\}\) indicate whether expert \(e\) in layer \(l\) is placed on GPU/rank \(g\). Each expert must be placed on exactly one rank:
\[
\sum_g x_{l,e,g} = 1, \quad \forall l,e.
\]
The load assigned to rank \(g\) in layer \(l\) is
\[
L_{l,g} = \sum_e B_{l,e} x_{l,e,g}.
\]
To balance expert load, Gimbal minimizes the deviation between each rank load and the average layer load:
\[
C_{\text{load}} =
\sum_l \sum_g \left(L_{l,g} - \bar{L}_l\right)^2.
\]
To capture source-aware communication, let \(D_{s,g}\) denote the communication cost between DP source \(s\) and the rank \(g\) hosting the selected expert. The source-aware communication cost is
\[
C_{\text{comm}} =
\sum_l \sum_s \sum_e \sum_g
A_{l,s,e} D_{s,g} x_{l,e,g}.
\]
Finally, Gimbal includes a migration cost to avoid excessive expert movement. Let \(x^0_{l,e,g}\) be the placement before rebalancing. The migration cost is
\[
C_{\text{mig}} =
\sum_l \sum_e \sum_g
M_{l,e,g} \left|x_{l,e,g} - x^0_{l,e,g}\right|,
\]
where \(M_{l,e,g}\) represents the cost of moving expert \(e\) in layer \(l\) to rank \(g\). The overall objective is
\[
\min_x
\quad
C_{\text{load}} + C_{\text{comm}} + C_{\text{mig}},
\]
subject to placement and capacity constraints. This formulation captures the desired trade-off: expert placement should balance total expert load, reduce source-dependent communication, and avoid unnecessary migration.

\subsection{Online Placement Workflow}

The above formulation provides a clear optimization objective; however, directly solving the full MINLP formulation online is computationally prohibitive for serving. In our experiments, solving the MINLP placement problem for a 48-layer Qwen3-30B MoE model takes roughly 15 seconds, which is unacceptable on the inference critical path. Therefore, Gimbal uses the MINLP formulation as an offline optimization reference and calibration target, and adopts a lightweight online heuristic for runtime placement.

At the end of each profiling window, Gimbal aggregates the expert-load matrix \(B\) and the source-aware matrix \(A\). The Expert Placement Manager first ranks experts by a hotness signal derived from aggregate expert load and source-aware activation volume. It then places experts greedily. For each hot expert, Gimbal enumerates feasible ranks that still have placement capacity and computes a local placement score:
\[
S_{l,e,g}
=
\alpha C^{\text{comm}}_{l,e,g}
+
\beta C^{\text{load}}_{l,e,g}
+
\gamma C^{\text{mig}}_{l,e,g}.
\]
Here, \(C^{\text{comm}}_{l,e,g}\) estimates the source-aware communication cost if expert \(e\) is placed on rank \(g\), \(C^{\text{load}}_{l,e,g}\) estimates the resulting rank-load imbalance, and \(C^{\text{mig}}_{l,e,g}\) estimates the migration cost relative to the current placement. The coefficients \(\alpha\), \(\beta\), and \(\gamma\) control the trade-off among communication reduction, load balance, and migration stability.

Gimbal assigns each expert to the feasible rank with the lowest score, breaking ties by preferring placements that avoid migration and then by choosing less-filled ranks. If the selected placement differs from the current one, Gimbal applies it through expert migration. The measured expert pressure from MoE execution is periodically fed back to the DP scheduler, closing the loop between expert placement and request dispatch.
Overall, this design avoids two extremes: balancing only aggregate expert load can ignore cross-DP communication, while optimizing only source locality can overload EP ranks or trigger excessive migration. By using a lightweight source-aware matrix and a calibrated online heuristic, Gimbal captures the main source-to-expert traffic structure without paying the cost of full online affinity profiling.

\section{Implementation}
\noindent\textbf{Prototype overview.}
We implement Gimbal on top of vLLM \cite{vllm}. The prototype replaces the default DP-engine selection logic and integrates source-aware expert profiling into the MoE/EPLB runtime, adding approximately 1.7K lines of Python and Triton code across the frontend scheduler, scheduler-statistics path, MoE dispatch path, and EPLB rearrangement runtime.

\noindent\textbf{Runtime trace path.}
Each backend engine reports a compact runtime trace through vLLM's asynchronous engine-statistics channel. Gimbal augments the original statistics with remaining prefill tokens, waiting prefill tokens, KV-cache usage, and backend MoE pressure. MoE pressure is implemented as a normalized token-equivalent expert load collected from EPLB expert-load counters, aggregated within the TP group of each DP engine, and fed back as a relative penalty in DP-engine selection. The frontend scheduler uses the latest available trace and applies lightweight compensation after each dispatch to account for newly assigned requests before the next trace update.

\noindent\textbf{MoE profiling path.}
Gimbal collects aggregate expert-load statistics and source-DP-to-expert statistics along the normal MoE dispatch path. The aggregate expert-load matrix reuses EPLB's expert-counting path and is collected over a rearrangement window used for placement updates, while the source-aware matrix is accumulated on GPU during routing and aggregated across EP ranks before rank-aware rearrangement.
To reduce overhead, Gimbal builds on PPLX communication kernels~\cite{pplx-kernels}, which we adopt due to our earlier vLLM version and limited GPU scale. Gimbal then reuses expert token-count information exposed by the all-to-all backend and uses a new Triton kernel to fuse source counting with expert mapping.

\noindent\textbf{Policy configuration.}
Gimbal uses a small set of policy parameters for DP-engine selection, local queue aging, and source-aware expert placement. For KV-cache protection, \emph{HighKV} (Alg.~\ref{alg:global-dp-engine-scheduler}) returns true when an engine's KV-cache usage exceeds 90\%. This follows the common practice of reserving most GPU memory for KV cache in open-source LLM serving frameworks and treats high KV usage as a hard pressure signal. \emph{LargeGap} (Alg.~\ref{alg:global-dp-engine-scheduler}) returns true when the KV-cache usage difference between the most- and least-loaded engines exceeds 10\%, which avoids reacting to small fluctuations while still detecting meaningful cross-engine memory imbalance. For local queue ordering, we set \(\theta_{age}\) (Alg.~\ref{alg:SJF with Aging}) to 5 seconds. In our testbed, the P99 TTFT under high load is below 4.9 seconds; therefore, a request waiting longer than 5 seconds is promoted to avoid starvation.
\begin{figure}
    \centering
    \includegraphics[width=1\columnwidth]{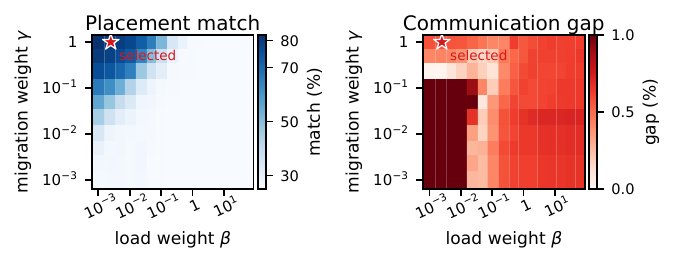}
        \vspace{-0.7cm}
    \caption{Calibration of Gimbal's online expert-placement heuristic against the offline MINLP reference.}
    \label{fig:eplb-weight-calibration}
\end{figure}
For the online expert-placement heuristic, Gimbal scores each candidate placement using source-aware communication cost, projected rank-load imbalance, and migration-stability cost. Since only the relative values of \(\alpha\), \(\beta\), and \(\gamma\) affect rank selection, we fix \(\alpha=1.0\) and calibrate \(\beta\) and \(\gamma\) against an offline MINLP reference computed from the dumped expert-load and source-DP-to-expert matrices. As shown in Figure~\ref{fig:eplb-weight-calibration}, the calibrated setting \((\alpha,\beta,\gamma)=(1.0,0.0025,1.0)\) preserves more than 80\% of the MINLP placement decisions and keeps source-aware communication within 0.6\% of the offline reference. We evaluate the performance impact of this calibration in Section~\ref{sec:impact-minlp-calibration}.

\begin{figure}
    \centering
    \includegraphics[width=1\columnwidth]{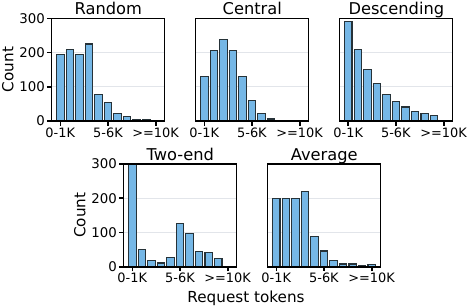}
        \vspace{-0.5cm}
    \caption{BurstGPT request-length distributions used in our evaluation.}
    \label{fig:burst-dataset-distribution}
    \centering
    \includegraphics[width=1\columnwidth]{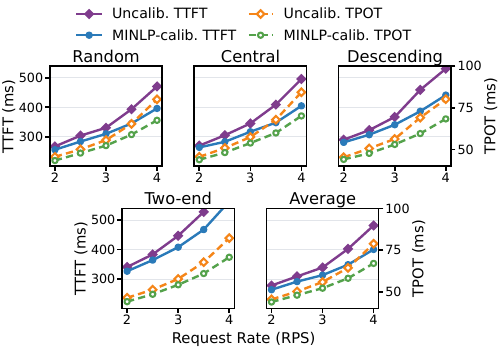}
    \vspace{-0.5cm}
    \caption{Impact of MINLP calibration on the greedy expert-placement policy across request rates under five BurstGPT workload distributions.}
    \label{fig:calibrated-greedy-comparison}
\end{figure}

\section{Evaluation}
\subsection{Experimental Setup}
We introduce our evaluation methodology in this section.

\noindent\textbf{Testbed.}
All experiments are conducted on a server with two 32-core Intel Xeon Platinum 8462Y+ CPUs, four NVIDIA HGX H100 80GB SXM5 GPUs connected by NVLink, and 1~TB of system memory. Following a common deployment paradigm~\cite{li2026semantic}, we use DP=2, TP=2, and EP=4: attention instances are replicated across two DP groups, attention computation within each DP group is sharded across two TP GPUs, and experts are partitioned across all four GPUs. Figure~\ref{fig:source-expert-comm} illustrates this deployment topology.

\noindent\textbf{Experimental Model.}
We evaluate Qwen3-30B-A3B~\cite{qwen3}, a representative medium-scale MoE model.

\noindent\textbf{Datasets and Traces.}
We use BurstGPT~\cite{burstgpt}, a real-world LLM serving trace widely used in prior work~\cite{zhang2025blitzscale,kim2025oaken}. Following recent studies~\cite{flexpipe,aegaeon}, we reshape the BurstGPT trace into five representative request-length distributions: \textit{Random}, \textit{Central}, \textit{Descending}, \textit{Two-end}, and \textit{Average}, as shown in Figure~\ref{fig:burst-dataset-distribution}. For each distribution, we sample 1,000 requests and repeat each distribution-rate experiment with three random seeds, reporting averaged results.

\noindent\textbf{Baselines and Our System.}
We compare Gimbal against three MoE serving baselines:
1) \textbf{vLLM}~\cite{vllm}, a widely adopted open-source LLM inference engine with rapid community development and broad use in academic systems. We use its default request-count-based DP scheduler together with its aggregate expert-activation-count-based EPLB mechanism as the production-grade baseline. We disable vLLM's built-in prefix-cache matching to reduce experimental bias.
2) \textbf{MoETuner}~\cite{go2025moetuner}, which optimizes MoE inference by exploiting expert affinity and offline traffic profiling for static expert placement. We implement its placement logic and affinity-based algorithm in our system and evaluate it under the same runtime and workload settings.
3) \textbf{Sem-MoE}~\cite{li2026semantic}, an MoE inference framework that includes static Integer Linear Programming (ILP) expert placement and a model-predicted semantic request-routing approach. As its semantic prediction model is unavailable to us, we implement an oracle variant by profiling the request-to-engine assignment in advance and replaying this mapping during evaluation, which assumes perfect knowledge of the semantic routing decision.


\noindent\textbf{Evaluation Metrics.}
We evaluate several key metrics commonly used in LLM serving:
\textit{TTFT}, measuring the latency from sending a request to receiving the first generated token;
\textit{TPOT}, representing the average decoding latency per output token, excluding the first generated token;
\textit{Throughput}, measuring the number of completed requests per second under a given load.

\begin{figure*}
    \centering
    \includegraphics[width=1\textwidth]{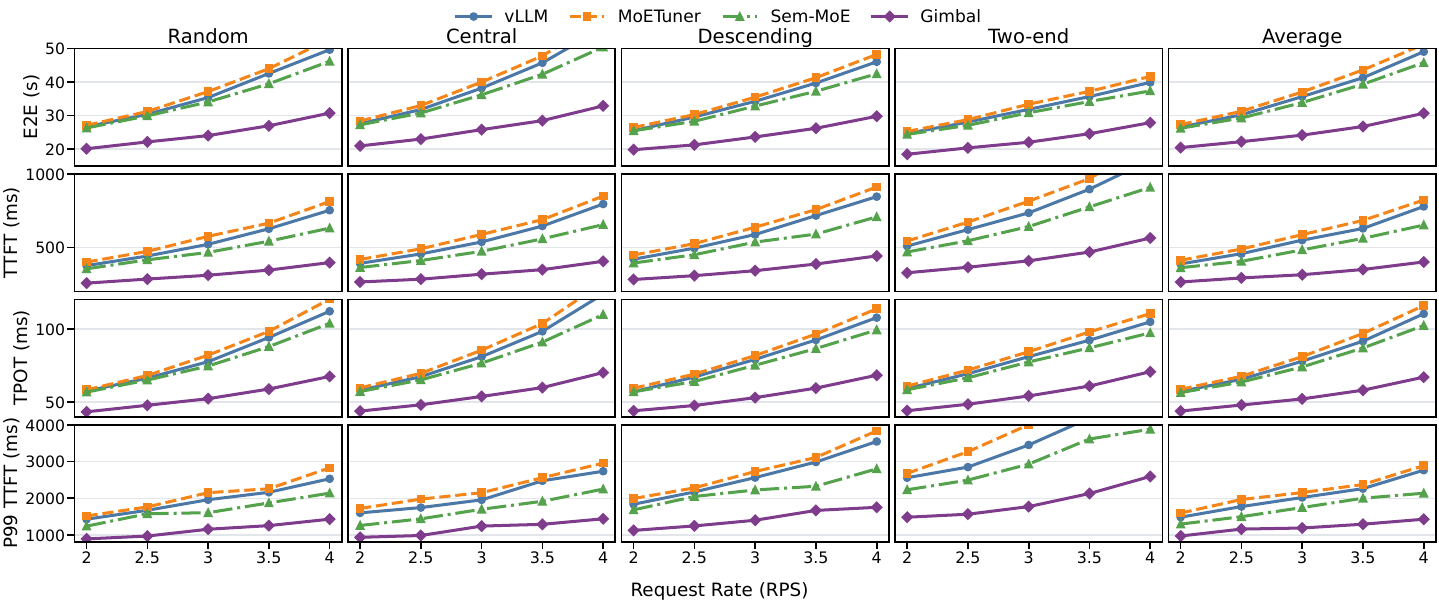}
            \vspace{-0.6cm}
    \caption{End-to-end latency, TTFT, TPOT, and P99 TTFT across request rates under five BurstGPT workload distributions. Each point averages three random seeds.}
                \vspace{-0.3cm}
    \label{fig:combined-latency-panel}
\end{figure*}

\subsection{Impact of MINLP Calibration}
\label{sec:impact-minlp-calibration}

We first evaluate the effect of calibrating the online greedy expert-placement policy against the offline MINLP reference. As discussed in Section~\ref{sec:source-aware-expert-lb}, the uncalibrated greedy policy places experts in descending hotness order and selects the locally best feasible rank according to a weighted score. However, when the weights are not calibrated, the policy can overreact to short-window load imbalance and move too many experts, disrupting source-aware locality.

Figure~\ref{fig:calibrated-greedy-comparison} compares the uncalibrated greedy policy with the MINLP-calibrated policy across five workload distributions and request rates. The calibrated policy consistently reduces both TTFT and TPOT. Across all request rates, workload distributions, and random seeds, MINLP calibration reduces average TTFT by 10.8\% and average TPOT by 9.2\% compared with the uncalibrated greedy policy. This improvement confirms that calibration helps preserve the desired trade-off among source-aware communication, load balance, and migration stability, instead of aggressively reshuffling experts based on transient load signals.
Therefore, in the remaining experiments, we use the MINLP-calibrated expert-placement heuristic as the default Gimbal configuration when comparing against other baselines.

\subsection{Performance}

\textbf{End-to-end latency.}
We first evaluate the overall serving latency of Gimbal and the baselines across all request rates and workload distributions. Figure~\ref{fig:combined-latency-panel} summarizes end-to-end latency, TTFT, TPOT, and P99 TTFT. Overall, Gimbal consistently achieves the lowest latency across the five BurstGPT distributions. Across all request rates, workload distributions, and random seeds, Gimbal reduces mean end-to-end latency by 32.0\% compared with vLLM, 34.7\% compared with MoETuner, and 28.5\% compared with Sem-MoE. The improvement becomes larger as the request rate increases: compared with vLLM, Gimbal reduces end-to-end latency by 23.8\% when request per second RPS=2 and by 36.9\% at RPS=4. This trend shows that Gimbal is most effective when the serving system is under higher pressure, where coarse request dispatching and static expert placement are more likely to amplify queueing delay and MoE execution imbalance.

\textbf{TTFT.}
Gimbal also significantly reduces TTFT, which is mainly affected by request dispatching, prefill work, and engine-side queueing. Across all settings, Gimbal reduces average TTFT by 42.9\% compared with vLLM, 47.0\% compared with MoETuner, and 34.7\% compared with Sem-MoE. The benefit again grows with load: compared with vLLM, the TTFT reduction increases from 33.1\% at RPS=2 to 48.3\% at RPS=4. To understand this behavior, we inspect the RPS=4 Random traces. vLLM's two engines appear balanced by request count, but they still carry high in-flight pressure: on average, the two engines keep about 91 and 90 running requests, with about 26--27\% KV-cache usage. Gimbal reduces this pressure to about 58 and 57 running requests, with about 17\% KV-cache usage. Moreover, the average prompt-throughput gap between the two engines decreases from 1.24K tokens/s in vLLM to 0.74K tokens/s in Gimbal. This indicates that request count alone does not capture token-side pressure, while Gimbal's fine-grained engine trace allows the frontend scheduler to avoid engines that are already busy in terms of prefill work, KV usage, or backend MoE pressure.

\textbf{TPOT.}
Gimbal reduces average TPOT by 33.3\% over vLLM, 36.2\% over MoETuner, and 29.5\% over Sem-MoE. Since TPOT is dominated by repeated decode iterations, this improvement comes from reducing MoE-layer execution and communication inefficiency. MoETuner and Sem-MoE improve expert placement through offline affinity or static routing information, but they cannot adapt to the online source-to-expert traffic generated by the request stream. In contrast, Gimbal collects source-DP-to-expert routing statistics during serving and uses them to guide expert placement. This allows Gimbal to reduce expert hotspots and source-dependent all-to-all traffic during decode. As load increases, the TPOT reduction over vLLM grows from 24.1\% at RPS=2 to 38.4\% at RPS=4, showing that online expert placement becomes increasingly important under heavier decode pressure.

\textbf{Throughput.}
We further examine whether the latency reduction comes at the cost of serving capacity. Figure~\ref{fig:throughput-combined} reports both normalized throughput across request rates and raw throughput at RPS=4. Gimbal achieves higher throughput than vLLM at every request rate: 0.94\% higher at RPS=2, 1.06\% at RPS=2.5, 1.52\% at RPS=3, 2.05\% at RPS=3.5, and 3.04\% at RPS=4. At RPS=4, Gimbal improves throughput over vLLM on all five distributions, with gains ranging from 1.26\% on Descending to 4.20\% on Random. These results show that Gimbal does not trade throughput for latency. Instead, by reducing avoidable queueing, lowering in-flight engine pressure, and improving MoE placement, Gimbal allows the system to complete requests faster under the same offered load.

Overall, Gimbal consistently outperforms all baselines across average and tail latency. Across all settings, it reduces mean TTFT/TPOT by 42.9\%/33.3\% over vLLM, 47.0\%/36.2\% over MoETuner, and 34.7\%/29.5\% over Sem-MoE. Furthermore, high-load throughput is increased by 3\%. As shown in Figure~\ref{fig:combined-latency-panel}, Gimbal also reduces P99 TTFT by 44.3\%, 48.6\%, and 33.8\% over the three baselines, respectively, confirming that coordinated cross-level scheduling is especially effective at reducing long-tail stalls under heterogeneous workloads.

\subsection{Ablation Experiment}
\begin{figure}[t]
    \centering
        \includegraphics[width=1\columnwidth]{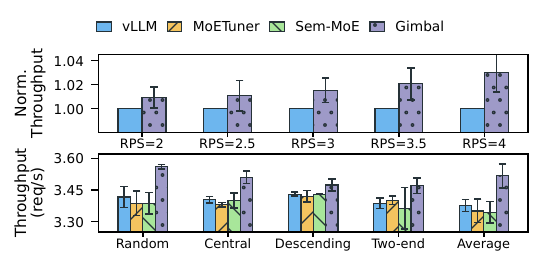}
    \vspace{-0.7cm}
        \caption{Request throughput comparison. Top: throughput normalized to vLLM at each request rate from RPS=2 to 4. Bottom: raw request throughput across five workload distributions at RPS=4.}
        \label{fig:throughput-combined}
    \centering
    \includegraphics[width=1\columnwidth]{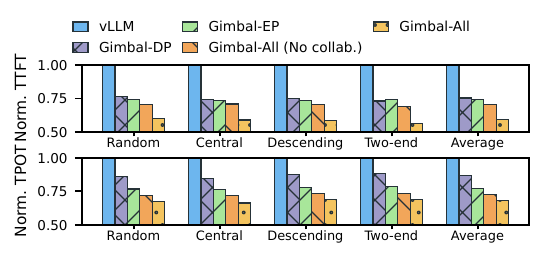}
    \vspace{-0.7cm}
    \caption{Normalized latency ablation averaged across request rates. Lower is better.}
    \label{fig:ablation-normalized-rps4}
    \centering
    \includegraphics[width=0.95\columnwidth]{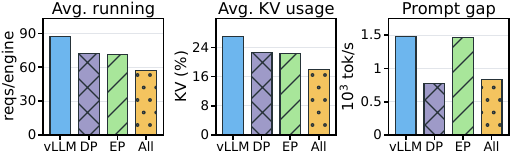}
    \vspace{-0.2cm}
    \caption{Runtime behavior under high load at RPS=4.}
    \label{fig:runtime-behavior-rps4}
\end{figure}

We conduct an ablation study to isolate the contribution of each component and, more importantly, the benefit of coordinated cross-level scheduling. We compare five configurations: 
\textbf{1) vLLM}, 
\textbf{2) Gimbal-DP}, which enables fine-grained DP-engine scheduling and local SJF-style ordering; 
\textbf{3) Gimbal-EP}, which enables source-aware expert load balancing; 
\textbf{4) Gimbal-All (No Collaboration)}, which enables both DP-side and EP-side optimizations but combines them independently without feeding backend MoE pressure into DP scheduling;
\textbf{5) Gimbal-All}, which enables the full coordinated design. 
Figure~\ref{fig:ablation-normalized-rps4} reports TTFT and TPOT normalized against vLLM, averaged across all request rates, workload distributions, and random seeds. 

Both individual components improve latency, but they target different bottlenecks. Gimbal-DP reduces normalized TTFT by 25.1\% and TPOT by 13.4\% by avoiding overloaded DP engines and reducing local queueing delay. Gimbal-EP reduces TTFT by 26.2\% and TPOT by 22.7\% by mitigating expert hotspots and improving source-aware expert placement. When both components are enabled without cross-level collaboration, Gimbal-All (No collab.) further reduces TTFT by 29.8\% and TPOT by 27.3\%. This confirms that the two optimizations are complementary, but also shows that simply adding them together does not fully exploit their interaction.

The full Gimbal design achieves the best result, reducing TTFT by 41.4\% and TPOT by 32.0\% over vLLM. Compared with Gimbal-All (No collab.), coordinated cross-level scheduling further reduces TTFT by 16.5\% and TPOT by 6.5\%. The key difference is that Gimbal-All feeds backend MoE pressure back to the DP scheduler, so request dispatching can react not only to frontend queue and KV-cache pressure, but also to expert-side pressure. This prevents the DP scheduler from sending more work to engines whose co-located expert ranks are under high MoE load.

We further investigate how Gimbal-DP, Gimbal-EP, and their coordinated combination improve serving latency. Figure~\ref{fig:runtime-behavior-rps4} reports three runtime signals collected from server traces at RPS=4: the average number of running requests per engine, the average KV-cache usage, and the cross-engine prompt-throughput gap. The prompt-throughput gap reflects how evenly prefill-token work is distributed across DP engines, while running requests and KV-cache usage capture the resulting engine-side pressure.
Gimbal-DP directly improves frontend dispatching by reducing token-side prefill imbalance: averaged across all RPS=4 runs, the cross-engine prompt-throughput gap drops from 1485.66 tokens/s in vLLM to 768.43 tokens/s. This indicates that prefill work is distributed more evenly across the two engines, avoiding the case where one engine is overloaded while the other remains underutilized. Meanwhile, Gimbal-EP reduces backend MoE pressure through source-aware expert placement, which indirectly accelerates decode execution and lowers the average number of running requests per engine from 87.56 in vLLM to 71.53. When combined, Gimbal-All coordinates these two effects: DP scheduling reduces prefill-side imbalance, while EP placement lowers decode-side MoE pressure. Together, by feeding backend MoE pressure back to the frontend scheduler, they jointly prevent expert-side hotspots from propagating into DP-engine imbalance, thereby reducing running requests and KV-cache usage and ultimately improving serving latency.

\subsection{System Overhead Analysis and Optimization}
\begin{figure}[t]
        \includegraphics[width=1\columnwidth]{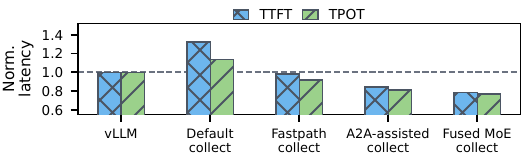}
    \vspace{-0.7cm}
        \caption{Normalized latency overhead of source-aware matrix collection. The no-collection vLLM serving path is normalized to 1.0.}
            \vspace{-0.7cm}
        \label{fig:source-aware-collection-cost}
\end{figure}

Since Gimbal collects expert activations and source-DP-to-expert mapping matrices online, this profiling path may introduce additional runtime overhead. We therefore analyze and optimize the overhead of matrix collection. Figure~\ref{fig:source-aware-collection-cost} shows the latency of collecting the source-DP-to-expert matrix. The default collection path introduces noticeable overhead because it records additional routing statistics outside the optimized MoE execution path. To reduce this cost, Gimbal applies three targeted optimizations. First, after analyzing the EPLB expert-statistics collection logic in vLLM, we add a fast collection path in the first EPLB profiling window and attach our collection logic to the existing EPLB execution path. Second, because Gimbal uses an all-to-all communication backend, it reuses token-count information exposed by the MoE communication backend, avoiding redundant expert-load accounting. Finally, Gimbal fuses the matrix collection logic into the normal inference path by implementing a dedicated Triton GPU kernel~\cite{tillet2019triton} that integrates source-aware counting into the regular MoE mapping path. The optimized path reduces collection overhead by eliminating extra mapping and accounting work, and even achieves lower latency than the original vLLM baseline.

\section{Related Work}

\textbf{Disaggregated LLM serving.}
Recent systems disaggregate inference phases to match their resource demands. Splitwise~\cite{splitwise}, DistServe~\cite{distserve}, and Mooncake~\cite{mooncake} separate prefill and decode execution, while TaiChi~\cite{taichi}, DOPD~\cite{dopd}, and PPD~\cite{ppd} further adapt the aggregation/disaggregation policy under different workloads and SLOs. Gimbal is complementary: instead of changing the prefill/decode deployment topology, it coordinates scheduling inside a standard DP+TP+EP MoE serving stack. Its pressure signals, such as KV-cache usage, remaining prefill work, and backend MoE pressure, could also serve as routing inputs in PD-disaggregated deployments.
\textbf{Attention/Expert Disaggregation.}
Another line of work separates computation by model component. JANUS~\cite{janus} disaggregates attention and MoE layers onto separate GPU pools, while Adrenaline~\cite{adrenaline} offloads part of decoding attention to prefill instances to improve utilization. These systems show that attention and MoE computation have different bottlenecks. Gimbal targets a different point: it keeps attention and experts co-located, and improves performance by coordinating DP-engine scheduling with source-aware expert placement. If attention and experts are physically disaggregated, Gimbal's feedback loop can be extended by feeding expert-pool pressure back to attention-side routing.

\noindent\textbf{Stage-level Scheduling and Heterogeneous Workloads.}
Stage-level disaggregation has also been explored for multimodal serving, where encode, prefill, and decode stages exhibit different compute and memory characteristics~\cite{hydrainfer,epdserve}. These works reinforce the need for scheduling policies that understand heterogeneous stages and dynamic workloads. Gimbal follows the same principle for MoE LLM serving, using lightweight online signals to react to bursty and long-context workloads where request length, KV-cache pressure, and expert routing skew change during serving.

\section{Conclusion}

We presented Gimbal, a coordinated cross-level scheduling system for MoE-based LLM serving. Gimbal coordinates frontend DP-engine scheduling with backend MoE expert placement by using fine-grained engine pressure, lightweight prefill-aware queue ordering, and online source-DP-to-expert routing statistics. Based on these signals, Gimbal dispatches requests away from overloaded engines and dynamically places experts to reduce expert hotspots and source-dependent communication cost. We prototyped Gimbal on top of vLLM and evaluated it on a 4-H100 testbed with Qwen3-30B-A3B and BurstGPT workloads. Experiments show that Gimbal reduces average TTFT by 42.9\% and average TPOT by 33.3\% compared with vLLM, while improving high-load request throughput by 3.0\%. 
Although evaluated on a single-node deployment, Gimbal's coordinated scheduling framework is not tied to Qwen-specific internals. Its mechanisms rely on general serving signals,
including KV-cache usage, prefill work, expert activation counts, and source-DP-to-expert routing statistics, which are commonly exposed by MoE serving systems. As communication and placement costs become increasingly important at scale, we expect cross-level coordination to provide greater benefits in large multi-node clusters. Future work includes evaluating Gimbal on additional MoE models, extending it to multi-node deployments, and exploring more adaptive expert-placement policies.
\bibliographystyle{ACM-Reference-Format}
\bibliography{sample-base}

\end{document}